\documentclass[conference]{IEEEtran}
\IEEEoverridecommandlockouts
\usepackage{cite}
\usepackage{amsmath,amssymb,amsfonts}
\usepackage{algorithmic}
\usepackage{caption}
\usepackage[percent]{overpic}
\usepackage{subcaption}
\usepackage{algorithmic}
\usepackage{algorithm}
\usepackage{graphicx}
\usepackage{textcomp}
\usepackage{xcolor}
\def\BibTeX{{\rm B\kern-.05em{\sc i\kern-.025em b}\kern-.08em
    T\kern-.1667em\lower.7ex\hbox{E}\kern-.125emX}}
\makeatletter
\def\endthebibliography{%
	\def\@noitemerr{\@latex@warning{Empty `thebibliography' environment}}%
	\endlist
}
\makeatother
\begin{document}

\title{Superimposed Channel Estimation in OTFS Modulation Using Compressive Sensing\\

}

\author{\IEEEauthorblockN{Omid Abbassi Aghda, Mohammad Javad Omidi}
\IEEEauthorblockA{\textit{Department of Electrical and Computer Engineering} \\
\textit{Isfahan University of Technology}\\
Isfahan 84156-83111, Iran \\
o.abasi@ec.iut.ac.ir, omidi@iut.ac.ir}
\and

\IEEEauthorblockN{Hamid Saeedi-Sourck}
\IEEEauthorblockA{\textit{Electrical Engineering} \\
\textit{Yazd University}\\
Yazd, Iran \\
saeedi@yazd.ac.ir}
}

\maketitle

\begin{abstract}
Orthogonal time frequency space (OTFS) technique is a two-dimensional modulation method that multiplexes information symbols in the delay-Doppler (DD) domain. OTFS combats high Doppler shift existing in high speed wireless communication. However, conventional channel estimation in OTFS suffers from high pilot overhead because guard symbols occupy a significant part of the DD domain grids. In this paper, a superimposed channel estimation is proposed which can completely estimate channel parameters without considering pilot overhead and performance degradation. As the channel state information (CSI) in the DD domain is sparse, a sparse recovery algorithm orthogonal matching pursuit (OMP) is used. Besides, our proposed method does not suffer from high peak to average power ratio (PAPR). To detect information symbols, a message passing (MP) detector, which exploits the sparsity of DD channel representation, is employed.
\end{abstract}


\section{Introduction}

High speed communication in the 6th generation of wireless communication is essential. However, there are some challenges to attain this goal, such as high Doppler shift, due to high speed, that causes frequency spreading. Frequency spreading leads to  inter carrier interference (ICI) in orthogonal frequency division multiplexing (OFDM). As a result, channel estimation and consequently data detection performance is not sufficient to provide highly reliable communication criteria. To overcome this issue, authors in \cite{Hadani2017} proposed a new modulation technique called orthogonal time frequency space (OTFS). OTFS system performance is not affected in the presence of high Doppler shift. But, there are still some challenges, such as channel estimation and data detection, that need attention \cite{Hadani2017,Raviteja2018a}.

OTFS modulation multiplexes information symbols in the delay-Doppler (DD) domain. Afterward, inverse discrete Zak transform (IDZT) is used to shape the domain signal. IDZT  spreads information symbols into different time slots. As a result, OTFS modulation can utilize the full diversity of the wireless channel \cite{Wei}. Moreover, the effect of each high speed terminal or object in the wireless channel are compensated separately. Hence, Doppler effect does not degrade the performance of the OTFS system \cite{Lampel2022,DDCbook}.

Channel estimation is an important task of every wireless system as data detection performance directly depends on estimation accuracy. DD representation of the wireless channel is sparse. It can be modeled just by a few taps and each tap is defined by three parameters; tap's coefficient, delay, and Doppler. Authors in \cite{Raviteja2019} proposed an embedded pilot (EP) scheme to estimate the DD channel state information (CSI). EP scheme detects the delay and Doppler parameters of each tap by setting a threshold on the DD domain and uses minimum mean square error (MMSE) to estimate taps gain. However, this method suffers from high pilot overhead as the pilot impulse in the DD domain needs to be separated from the data information to avoid pilot contamination. Moreover, EP scheme might lead to high peak to average power ratio (PAPR). To overcome the pilot overhead in the EP scheme, the authors in \cite{Mishra2021} proposed superimposed channel estimation where each pilot sequence is superimposed to the information symbols. This method leads to  the information data interference into the pilot symbols at the receiver. Therefore, an iterative method is used in \cite{Mishra2021} to cancel out data interference. This method uses MMSE method to estimate the channel vector, but it is assumed that the receiver knows the exact delay and Doppler of each tap. MMSE Superimposed pilot estimation considers an EP frame in the beginning of each super-frame to find out the delay and Doppler of each tap. So, it is assumed that the delay and Doppler of each tap do not change during a super-frame. A super-frame is considered to be made of multiple OTFS frames. Hence, the pilot overhead is not completely removed. To reduce the complexity of the superimposed channel estimation, the authors in \cite{Yuan2021} proposed superimposed channel estimation, where a single pilot is superimposed on the DD domain. However, single pilot superimposed estimation suffers from high PAPR and low estimation precision. Therefore, there is a need to find out a method that estimates the delay and Doppler of each tap without imposing high PAPR and overhead.

Detection in the OTFS system is considered the most challenging part of the OTFS system as it requires high computational complexity. The authors in \cite{Raviteja2018a} proposed a low-complexity message passing (MP) detector which is designed based on the sparse nature of the DD representation of the wireless channel. This method can efficiently cancel out the inter symbols interference (ISI) and ICI of the received signal. An MP detector is used in this paper for data detection.

In this paper, similar to \cite{Mishra2021}, a joint channel estimation and data detection method is proposed, where each pilot symbol is superimposed on the data symbol. However, as our estimation is based on the sparse recovery algorithm, the proposed method can detect the delay and Doppler of each tap, unlike the one in \cite{Mishra2021}. Therefore, there is no need to consider a dedicated frame with an embedded estimation method for the following frames which reduces pilot overhead. Moreover, OMP superimposed method similar to MMSE superimposed has much less PAPR compared to the EP scheme and single pilot superimposed estimation method.

The rest of the paper is organized as follows. In Section-II, the system model of the OTFS system is analyzed. In Section-III, proposed method to estimate the channel parameters and detect data is provided. In  Section-IV simulation results are illustrated and fully explained. This is followed by conclusion in Section-V.

In this paper, $\mathbf{A}$, $\mathbf{a}$, and $a$ are used to show matrices, vectors, and scalar values. Moreover, the notation $(.)^H$ shows Hermitian transposition operator and $\mathbf{A}\otimes\mathbf{B}$	is Kronecker product of matrices $\mathbf{A}$ and $\mathbf{B}$. To vectorize matrix $\mathbf{A}$ column-wise, $vec\left(\mathbf{A}\right)$ and to do inverse $vec^{-1}\left(\mathbf{A}\right)$ is used. Also, $\mathbf{F}_M$ indicates M-point discrete Fourier transform (DFT) matrix.

 \begin{figure*}[!th]
 	\centering
 	\begin{overpic}[width=\linewidth,tics=10]{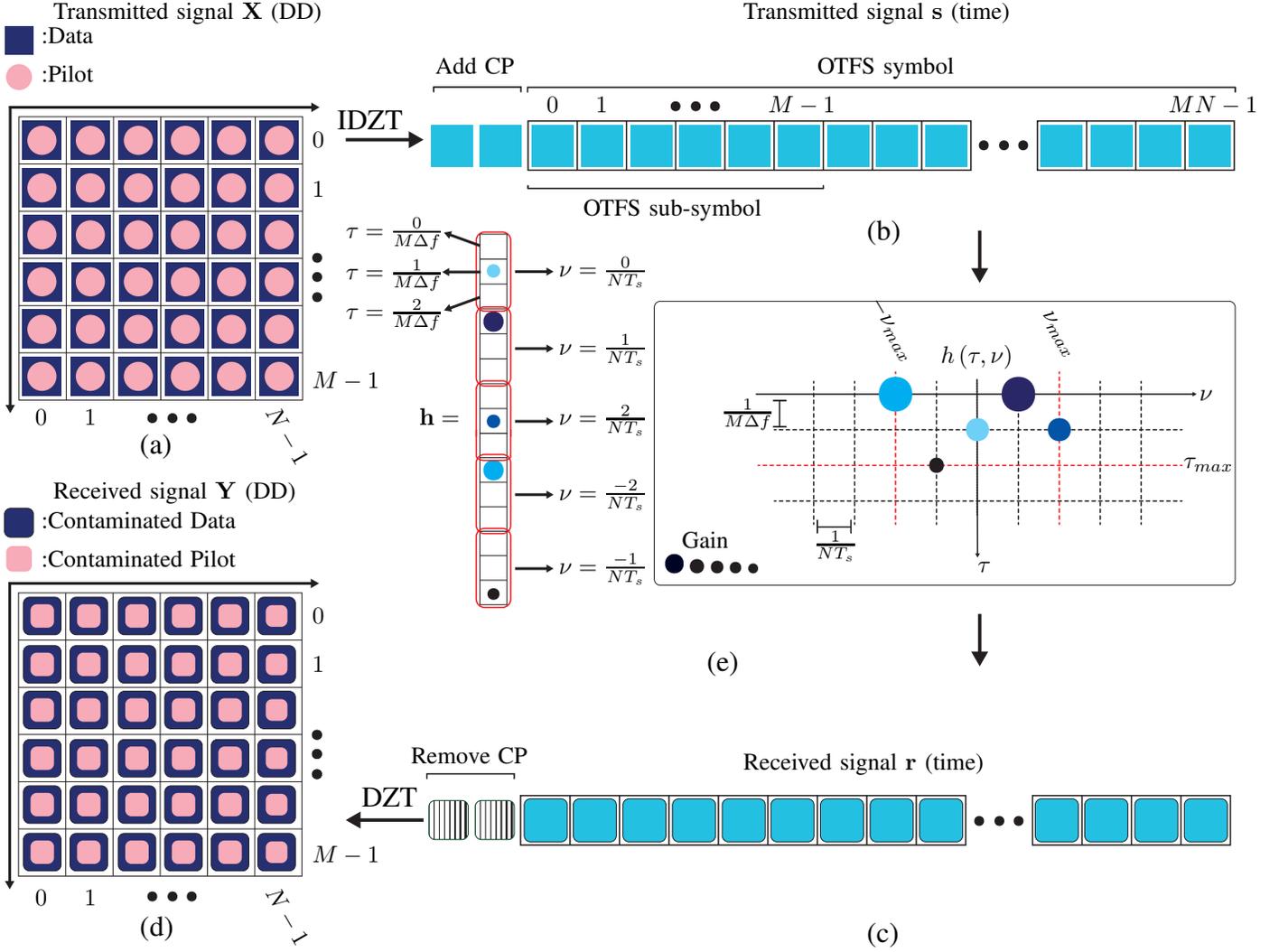}
 		\put (27,67) {\scalebox{1.2}{IDZT} }
 		\put (29,12) {\scalebox{1.2}{DZT} }
 		\put (11,40.7) {\scalebox{1.2}{(a)} }
 		\put (70,58) {\scalebox{1.2}{(b)}}
 		\put (57,23) {\scalebox{1.2}{(e)}  }
 		\put (11,1.5) {\scalebox{1.2}{(d)}  }
 		\put (70,1) {\scalebox{1.2}{(c)}  }
 		\put (3,70.9) {\scalebox{1}{:Pilot}  }
 		\put (3,74) {\scalebox{1}{:Data}  }
 		\put (3,31.5) {\scalebox{1}{:Contaminated Pilot}  }		
 		\put (3,34.6) {\scalebox{1}{:Contaminated Data}  }
 		\put (25,65.5) {\scalebox{1}{$0$}  }
 		\put (25,61.5) {\scalebox{1}{$1$}  }		
 		\put (25,46) {\scalebox{1}{$M-1$}  }		
 		\put (2.5,43) {\scalebox{1}{$0$}  }
 		\put (6.5,43) {\scalebox{1}{$1$}  }		
 		\put (21,44) {\scalebox{1}{\rotatebox{-60}{$N-1$}}  }		
 		\put (25,26.9) {\scalebox{1}{$0$}  }
 		\put (25,22.9) {\scalebox{1}{$1$}  }		
 		\put (25,7.5) {\scalebox{1}{$M-1$}  }		
 		\put (2.5,4) {\scalebox{1}{$0$}  }
 		\put (6.5,4) {\scalebox{1}{$1$}  }		
 		\put (21,5) {\scalebox{1}{\rotatebox{-60}{$N-1$}}  }					
 		\put (4,76) {\scalebox{1}{Transmitted signal $\mathbf{X}$ (DD)}  }
 		\put (4,37) {\scalebox{1}{Received signal $\mathbf{Y}$ (DD)}  }		
 		\put (76,48) {\scalebox{1}{$h\left(\tau,\nu\right)$}}			
 		\put (79,31) {\scalebox{1}{$\tau$}}					
 		\put (97,45) {\scalebox{1}{$\nu$}}					
 		\put (55,33) {\scalebox{1}{Gain}}		
 		\put (70,53) {\scalebox{1}{\rotatebox{-60}{$-\nu_{max}$}}}									
 		\put (84,52) {\scalebox{1}{\rotatebox{-60}{$\nu_{max}$}}}									
 		\put (95.7,39.5) {\scalebox{1}{$\tau_{max}$}}											
 		\put (58.2,43.5) {\scalebox{1}{$\frac{1}{M\Delta f}$}}													
 		\put (65.7,32.8) {\scalebox{1}{$\frac{1}{NT_s}$}}
 		\put (60,76) {\scalebox{1}{Transmitted signal $\mathbf{s}$ (time)}}
 		\put (60,15) {\scalebox{1}{Received signal $\mathbf{r}$ (time)}}
 		\put (33,43) {\scalebox{1}{ $\mathbf{h}=$ }}
 		\put (47,60) {\scalebox{1}{OTFS sub-symbol}}
 		\put (66,71.5) {\scalebox{1}{OTFS symbol}}
 		\put (45,36.9) {\scalebox{1}{$\nu = \frac{-2}{NT_s}$}}
 		\put (45,30.9) {\scalebox{1}{$\nu = \frac{-1}{NT_s}$}}
 		\put (45,48.7) {\scalebox{1}{$\nu = \frac{1}{NT_s}$}}	
 		\put (45,43) {\scalebox{1}{$\nu = \frac{2}{NT_s}$}}					
 		\put (45,55) {\scalebox{1}{$\nu = \frac{0}{NT_s}$}}							
 		\put (35,71.5) {\scalebox{1}{Add CP}}					
 		\put (33,15.5) {\scalebox{1}{Remove CP}}							
 		\put (27.7,54.8) {\scalebox{1}{$\tau = \frac{1}{M\Delta f}$}}	
 		\put (27.7,51.6) {\scalebox{1}{$\tau = \frac{2}{M\Delta f}$}}			
 		\put (27.7,58.2) {\scalebox{1}{$\tau = \frac{0}{M\Delta f}$}}	
 		\put (44,68.4) {\scalebox{1}{$0$}}					
 		\put (48,68.4) {\scalebox{1}{$1$}}					
		\put (62,68.4) {\scalebox{1}{$M-1$}}					 		
		\put (94.5,68.4) {\scalebox{1}{$MN-1$}}					 		
 	\end{overpic}
 	\caption{Schematic of the superimposed channel estimation in the OTFS system: (a) DD representation of the transmitted signal; (b) time domain representation of the transmitted signal; (c) time domain representation of the received signal; (d) DD representation of the received signal; (e) DD representation of the $h\left(\tau,\nu\right)$ (right hand figure) and channel vector $\mathbf{h}$ (left hand figure)}.
 	\label{fig:DD_data_pilot_transmitter}
 \end{figure*}
\section{System Model}

In the OTFS system, information symbols and pilot sequences are arranged in the DD domain. Then, the time domain signal is shaped using IDZT and transmitted into the time-varying (TV) channel. At the receiver, the time domain signal is transformed into the DD domain using discrete Zak transform (DZT) \cite{Thaj2020}, and channel estimation and data detection are performed.
\subsection{OTFS Modulation}
In the proposed method, information quadrature amplitude modulation (QAM) symbols and pilot sequences are superimposed in the DD domain at the transmitter, Fig. 1(a). The DD domain grid is divided into $M$ and $N$ along delay and Doppler axis respectively. Correspondingly, no pilot overhead is considered for the system, unlike conventional estimation methods proposed and classified in \cite{Raviteja2019,Otfs2019} and \cite{Naikoti2021}. The DD representation of the transmitted signal is $\mathbf{X} = \mathbf{X}_d + \mathbf{X}_p \in\mathbb{C}^{M\times N}$, where $\mathbf{X}_d$ and $\mathbf{X}_p$ are the DD domain matrix for data information and pilot, respectively. Its elements $X\left[k,l\right] =X_d\left[k,l\right]+X_p\left[k,l\right] $ contains superimposed pilots on data symbols. It is important to be mentioned that the average power of the transmitted signal is equal to 1; thus $\sigma^2=\sigma^2_d+\sigma^2_p = 1$. It is shown in \cite{Mishra2021} that  setting the average pilot power to $\sigma^2_p=0.3$,  equal to 30$\%$ of the total power, guarantee bit error rate (BER) and normalize mean square error (NMSE) for the data detection and channel estimation, respectively. Therefore, average data power is assumed to be $\sigma^2_d=0.7$ in this work. Pilot sequences are generated from Gaussian distribution $\mathcal{N}\left(0,\,\sigma^{2}_p\right)$.

Using IDZT, the time domain signal is shaped, Fig. 1(b). IDZT transforms a 2D signal from the DD domain into a 1D signal in the time domain \cite{Bolcskei1996}. Using IDZT, the time domain signal $\mathbf{s}\in\mathbb{C}^{MN\times1} $ is equal to \cite{Thaj2020,Raviteja2019a}
\begin{equation}
\mathbf{s}=IDZT\left(\mathbf{X}\right) = vec\left(\mathbf{XF}_N^H\right)=(\mathbf{F}_N^H\otimes\mathbf{I}_M)\mathbf{x},
\label{eq:idzt of DD signal in transmitter}
\end{equation}
where $\mathbf{x} = vec\left(\mathbf{X}\right)\in\mathbb{C}^{MN\times 1}$.

Before feeding the discrete time domain signal to digital to analog converter (DAC) in order to form the continuous time signal $r\left(t\right)$, cyclic prefix (CP) is added to the beginning of the OTFS symbol to avoid interference from previous OTFS symbol, Fig. 1(b). Clearly, CP length must be at least equal to the maximum delay spread of the wireless channel \cite{DDCbook}.

\subsection{OTFS Demodulation}
At the receiver, the continuous time domain signal $r\left(t\right)$ is received and using analog to digital converter (ADC), the received discrete time domain vector $\mathbf{r}$ can be made Fig. 1(c). After removing CP, the inverse operation of the transmitter is applied to the vector $\mathbf{r}$ to obtain the received DD domain matrix $\mathbf{Y}\in\mathbb{C}^{M\times N}$, as shown in Fig. 1(d). Thus, $\mathbf{Y}$ can be written as 
\begin{equation}
	\mathbf{Y} = vec^{-1}\left(\mathbf{r}\right)\mathbf{F}_N.
	\label{eq:vectorized_DZT}
\end{equation}
In fact \eqref{eq:vectorized_DZT} represents DZT of the vector $\mathbf{r}$. The inverse operation of \eqref{eq:vectorized_DZT}, that is IDZT of $\mathbf{Y}$ to obtain $\mathbf{r}$, can be expressed as  \cite{Raviteja2019a}
\begin{equation}
 \mathbf{r}= IDZT\left(\mathbf{Y}\right) = vec\left(\mathbf{YF}_N^H\right)=(\mathbf{F}_N^H\otimes\mathbf{I}_M)\mathbf{y} ,
	\label{eq:dzt_of_received_signal}
\end{equation}
where $\mathbf{y} = vec\left(\mathbf{Y}\right)\in\mathbb{C}^{MN\times 1}$

\subsection{DD Channel Model}
The received signal over the doubly selective channel is denoted as $r\left(t\right)$ and is given by \cite{Hadani2017}
\begin{equation}
	r\left(t\right) = \int\int h\left(\tau,\nu\right)s\left(t-\tau\right)e^{j2\pi\nu\left(t-\tau\right)}d\tau d\nu + w\left(t\right).
	\label{eq:channel_effect_continuous_time_domain}
\end{equation}
In \eqref{eq:channel_effect_continuous_time_domain}, $s\left(t\right)$ is the transmitted signal and  $h\left(\tau,\nu\right)$ denotes the DD representation of doubly-selective channel. Fig. 1(e) (rigth hand figure) depicts the DD domain CSI.  $h\left(\tau,\nu\right)$ is given by \cite{Raviteja2019a}
\begin{equation}
	h\left(\tau,\nu\right) = \sum_{i=1}^{P}h_i\delta\left(\tau-\tau_i,\nu-\nu_i\right).
	\label{eq:DD_channel_representation}
\end{equation}
$h\left(\tau,\nu\right)$ does not change during the DD coherence time which means the transmitted signal experience constant channel during a long frame \cite{Thaj2020,Srivastava2022}. The feature that $h\left(\tau,\nu\right)$ does not change during a long frame, comes from the fact that \eqref{eq:DD_channel_representation} represents a meaningful physical picture of the channel that is the distance and velocity of each object does not change for a few milliseconds \cite{Wiffen2018,Srivastava2022}.
According to \eqref{eq:DD_channel_representation}, $h\left(\tau,\nu\right)$ has sparse representation \cite{Deka2021}. Therefore, channel estimation is performed using sparse recovery algorithms  \cite{Paper2013}, which is applied in this work.

Let us assume that the total frame bandwidth and duration are $B$ and $T_s$, respectively. Hence, $\Delta f = \frac{B}{M}$ is the sub-carrier spacing and $T = \frac{T_s}{N}$ is the duration of every OTFS sub-symbol which is equivalent to OFDM symbol duration, in such a manner that the equation $T\Delta f=1$ holds. Delay and Doppler shift corresponding to the $i$th tap out of $P$ taps for $h\left(\tau,\nu\right)$ can be expressed as $\tau_i = \frac{l_i}{M\Delta f}$ and $\nu_i=\frac{k_i}{NT}$, respectively. $l_i$ and $k_i$ are assumed to be integer in our case, since this work does not consider fractional delay and Doppler as a result of low-resolution sampling in DD domain. It should be noted that only fractional Doppler might be considered in practical cases since the delay resolution is sufficient enough to consider $l_i$ as integer \cite{Raviteja2018a}.

By substituting \eqref{eq:DD_channel_representation} in \eqref{eq:channel_effect_continuous_time_domain}, the received signal after sampling and removing CP can be written as 
\begin{equation}
	r\left(n\right) = \sum_{i=1}^{P}h_ie^{j2\pi \frac{k_i\left(n-l_i\right)}{MN}}s\left(\left[n-l_i\right]_{MN}\right) + w\left(n\right),
	\label{eq:discrete_time_channel_effect_sigma}
\end{equation}
where $\left[.\right]_N$ is modulo $N$ operator and $w\left(n\right)$ is i.i.d additive white Gaussian noise (AWGN) with variance $\sigma_n^2$.
 The equation \eqref{eq:discrete_time_channel_effect_sigma}  is shown in matrix form as \cite{Srivastava2022},
\begin{equation}
	\mathbf{r} = \mathbf{H}\mathbf{s} + \mathbf{w} = \mathbf{H}\mathbf{s}_d+\mathbf{H}\mathbf{s}_p + \mathbf{w},
	\label{eq:channel_effect_vector_time_domain}
\end{equation}
where $\mathbf{H}\in\mathbb{C}^{MN\times MN}$ is the time domain channel matrix.
By substituting  \eqref{eq:idzt of DD signal in transmitter} and \eqref{eq:dzt_of_received_signal} in \eqref{eq:channel_effect_vector_time_domain}, it can be shown that vectors $\mathbf{y}$ and $\mathbf{x}$ are connected via the effective channel matrix $\mathbf{H}_{eff}\in\mathbb{C}^{MN\times MN}$. Therefore, the effective input-output relation of OTFS system is given by
\begin{equation}
	\mathbf{y} = \mathbf{H}_{eff}\mathbf{x} + \mathbf{\tilde{w}}=\mathbf{H}_{eff}\mathbf{x}_d+\mathbf{H}_{eff}\mathbf{x}_p+\mathbf{\tilde{w}},
	\label{eq:effective_input_output_relation}
\end{equation}
where $\mathbf{\tilde{w}} = \left(\mathbf{F}_N\otimes\mathbf{I}_M\right)\mathbf{w}$ \cite{Srivastava2022}. The relation between $\mathbf{H}_{eff}$ and $\mathbf{H}$ is straightforward as \cite{Raviteja2019a}
\begin{equation}
	\mathbf{H}_{eff} = \left(\mathbf{F}_N\otimes\mathbf{I}_M\right)\mathbf{H}\left(\mathbf{F}^H_N\otimes\mathbf{I}_M\right),
\end{equation}
where $\mathbf{H}_{eff}$ is a block-circulant matrix which its blocks are circulant. Moreover, $\mathbf{H}$ a is circulant matrix \cite{Thaj2020}.
\section{ Channel Estimation And Data Detection}
In this section, we explain the proposed method to estimate channel parameters and detect data jointly. As the CSI representation in the DD domain is sparse, to estimate channel parameters, the sparse recovery algorithm OMP is used. Afterward, the MP detector, which can exploit the sparsity of $\mathbf{H}_{eff}$,  detects data  . Since, pilots and data are superimposed, to estimate CSI and detect data perfectly an iterative interference cancellation is used. Therefore, the estimation and detection is categorized in coarse and fine step.
\subsection{Coarse Channel Estimation And Data Detection}
To estimate channel parameters, \eqref{eq:effective_input_output_relation} is reformulated. Therefore, it can be written as 
\begin{equation}
	\mathbf{y} = \mathbf{X'}\mathbf{h} + \mathbf{\tilde{w}} = \mathbf{X'}_d\mathbf{h} + \mathbf{X'}_p\mathbf{h} + \mathbf{\tilde{w}},
	\label{eq:effective_input_output_estimateshape}
\end{equation}
where $\mathbf{X'} = \mathbf{X'}_d + \mathbf{X'}_p \in\mathbb{C}^{MN\times (2k_{max}+1)(l_{max}+1)} $ and $\mathbf{h}\in\mathbb{C}^{(2k_{max}+1)(l_{max}+1)\times 1}$ is the channel vector \cite{Zhao2020}. In fig. 1(e) (left hand figure) it is shown how to form channel vector $\mathbf{h}$. is  As it is clear in \eqref{eq:effective_input_output_estimateshape}, to estimate channel vector $\mathbf{h}$, data interference into the pilot is considered as noise. Thus, \eqref{eq:effective_input_output_estimateshape} is rewritten as 
\begin{equation}
	\mathbf{y} =  \mathbf{X'}_p\mathbf{h} + \mathbf{\bar{w}},
	\label{eq:effective_input_output_estimateshape_inter}
\end{equation}
where $\mathbf{\bar{w}} =\mathbf{X'}_d\mathbf{h} +  \mathbf{\tilde{w}} $ is the interference into the pilot.
 By considering that the receiver knows the sparsity order of the vector $\mathbf{h}$, that is $\frac{P}{(2k_{max}+1)(l_{max}+1) }$, the channel vector can be estimated coarsely from \eqref{eq:effective_input_output_estimateshape_inter} using OMP algorithm  \cite{Paper2013}. The estimated channel vector in the coarse channel estimation is shown by $\mathbf{\hat{h}}^1$ where it can be considered as the $(1)^{st}$ iteration of the proposed iterative channel estimation. Now, using the estimated CSI, the data detection is performed.

To detect information data in the OTFS system, MP algorithm is proposed in \cite{Raviteja2018a} which is designed according to the sparse structure of the $\mathbf{H}_{eff}$. Coarse detected data in this step is called $\mathbf{\hat{x}}_d^1$. However, before the detection procedure, the pilot interference must be canceled out from the received data. Using $\mathbf{\hat{h}}^1$, effective estimated channel matrix $\mathbf{\hat{H}}_{eff}^1$ can be formed \cite{Raviteja2019a}. By subtracting contaminated pilot from the received DD vector $\mathbf{y}$ in \eqref{eq:effective_input_output_relation}, it can be seen that
\begin{equation}
	\mathbf{y}_d^1 = 	\mathbf{y} - \mathbf{\hat{H}}_{eff}^1\mathbf{x}_p =\mathbf{H}_{eff}\mathbf{x}_d+\left(\mathbf{H}_{eff}-\mathbf{\hat{H}}_{eff}^1\right)\mathbf{x}_p+\mathbf{\tilde{w}}. 
	\label{eq:cancel_pilot_interference}
\end{equation}
By considering $\mathbf{w}^{1}=\left(\mathbf{H}_{eff}-\mathbf{\hat{H}}_{eff}^1\right)\mathbf{x}_p+\mathbf{\tilde{w}}$ as interference to the received data, \eqref{eq:cancel_pilot_interference} can be shortened to
\begin{equation}
	\mathbf{y}_d^1 =\mathbf{H}_{eff}\mathbf{x}_d + \mathbf{w}^1.
	\label{eq:effective_realation_formp}
\end{equation}
$\mathbf{y}_d^1$ is fed to MP detector to detect $\mathbf{x}_d^1$. It is clear that due to the high power interference term in \eqref{eq:effective_input_output_estimateshape_inter}, the estimated CSI is not good enough to cancel out pilot interference and detect data with low error rate. Therefore, to refine channel estimation and consequently cancel out pilot interference, the detected data in the $(i)^{th}$ iteration helps the channel estimation in the $(i+1)^{th}$ iteration.

\subsection{Fine Channel Estimation And Data Detection}
Here, we use the detected data $\mathbf{\hat{x}}_d^i$ in the previous iteration as an aid to the pilot sequences for better channel estimation in the $(i+1)^{th}$ iteration. So, by rewriting \eqref{eq:effective_input_output_estimateshape_inter}, we have \cite{Mishra2021}.
\begin{align}
\mathbf{y}&= \left(\mathbf{X'}_p+\mathbf{\hat{X'}}_d^i\right)\mathbf{h} +\left(\mathbf{X'}_d-\mathbf{\hat{X'}}_d^i\right)\mathbf{h} + \mathbf{\tilde{w}}\nonumber \\
 &= \left(\mathbf{X'}_p+\mathbf{\hat{X'}}_d^i\right)\mathbf{h} + \mathbf{w}_d,
	\label{eq:reformulated_effective_input_output_for_iter_estimation}
\end{align}
where $\mathbf{w}_d = \left(\mathbf{X'}_d-\mathbf{\hat{X'}}_d^i\right)\mathbf{h} + \mathbf{\tilde{w}}$. It can clearly be seen from \eqref{eq:reformulated_effective_input_output_for_iter_estimation} that the term $\mathbf{\hat{X'}}_d^i$ is considered as a pilot to increase overall pilot power. So, this method is considered as joint channel estimation and data detection. Therefore, the better detection in the previous iteration may results in the better channel estimation in the current iteration. The channel vector is estimated using OMP algorithm  and shown by $\mathbf{\hat{h}}^{i+1}$ that is for $(i+1)^{th}$th iteration.

To detect data in the $(i+1)^{th}$ iteration, $\mathbf{\hat{h}}^{i+1}$ is used in order to detect data. However, before detection in the $(i+1)^{th}$ iteration, it is required to cancel out the pilot interference from received data. Interference cancellation is written  as
\begin{equation}
\mathbf{y}_d^{i+1} = 	\mathbf{y} - \mathbf{\hat{H}}_{eff}^{i+1}\mathbf{x}_p =\mathbf{H}_{eff}\mathbf{x}_d+\left(\mathbf{H}_{eff}-\mathbf{\hat{H}}_{eff}^{i+1}\right)\mathbf{x}_p+\mathbf{\tilde{w}}. 
\label{eq:cancel_pilot_interference_iterative}
\end{equation}
Similar to the $(i)^{th}$ data detection, here the interference term is named $\mathbf{w}^{i+1}=\left(\mathbf{H}_{eff}-\mathbf{\hat{H}}_{eff}^{i+1}\right)\mathbf{x}_p+\mathbf{\tilde{w}}$. $\mathbf{y}_d^{i+1}$ is fed to the MP detector to detect the data vector and the detected data in the $(i+1)^{th}$ iteration is called $\mathbf{\hat{x}}_d^{i+1}$. Algorithm 1 summarizes the joint channel estimation and data detection using OMP.

 It should be mentioned that as MP detector needs to know the variance of the interference and noise power, the covariance matrix of the channel vector is assumed to be known at the receiver.  The authors in \cite{Mishra2021}, calculated the interference power in detail.

 \begin{algorithm}
	\caption{Iterative channel estimation and data detection}
	\begin{algorithmic}[1]
		\renewcommand{\algorithmicrequire}{\textbf{Input:}}
		\renewcommand{\algorithmicensure}{\textbf{Output:}}
		\REQUIRE Pilot matrix $\mathbf{X}'_p$, received DD vector $\mathbf{y}$
		\ENSURE  Estimated channel vector $\mathbf{\hat{h}}$ and detected data vector $\mathbf{\hat{x}_d}$
		\\ \textit{Initialization} : $noi=$number of iteration
		\STATE estimate channel vector $\mathbf{\hat{h}}^1$ from \eqref{eq:effective_input_output_estimateshape_inter} using OMP
		\STATE cancel pilot interference from the received data using \eqref{eq:cancel_pilot_interference}
		\STATE detect data vector $\mathbf{\hat{x}}_d^1$ using MP algorithm from $\mathbf{y}_d^1$
		\FOR {$i = 2$ to $noi$}
		\STATE Update $\mathbf{X'}_p=\mathbf{X'}_p+\mathbf{\hat{X'}}_d^{i-1}$ 
		\STATE estimate channel vector $\mathbf{\hat{h}}^i$ from \eqref{eq:reformulated_effective_input_output_for_iter_estimation} using OMP
		\STATE cancel pilot interference from the received data using \eqref{eq:cancel_pilot_interference_iterative}
		\STATE detect data vector $\mathbf{\hat{x}}_d^i$ using MP algorithm from $\mathbf{y}_d^i$
		\ENDFOR
		\RETURN $\mathbf{\hat{h}}^{noi },\mathbf{\hat{x}}_d^{noi}$ 
	\end{algorithmic} 
\end{algorithm}

\section{Simulation Results}
\begin{figure}[!t]
	\centering
	\includegraphics[width=1\linewidth]{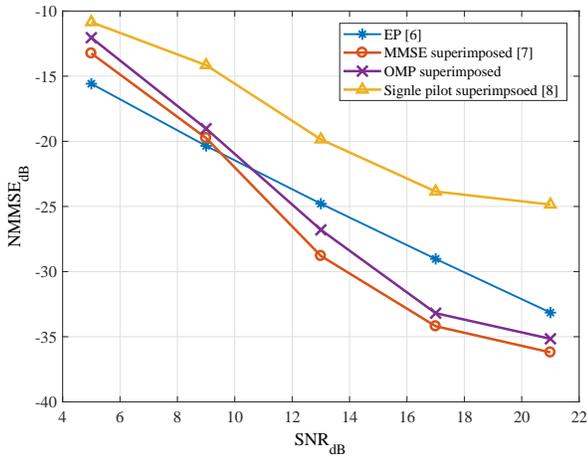}
	\caption{NMMSE of CSI estimation versus SNR.}
	\label{nmmse}
\end{figure}

\begin{figure}[!t]
	\centering
	\includegraphics[width=1\linewidth]{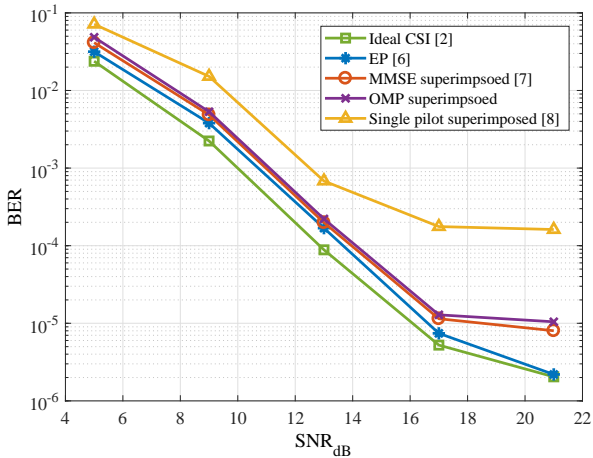}
	\caption{BER versus SNR.}
	\label{ber}
\end{figure}
\begin{figure}[!t]
	\centering
	\includegraphics[width=1\linewidth]{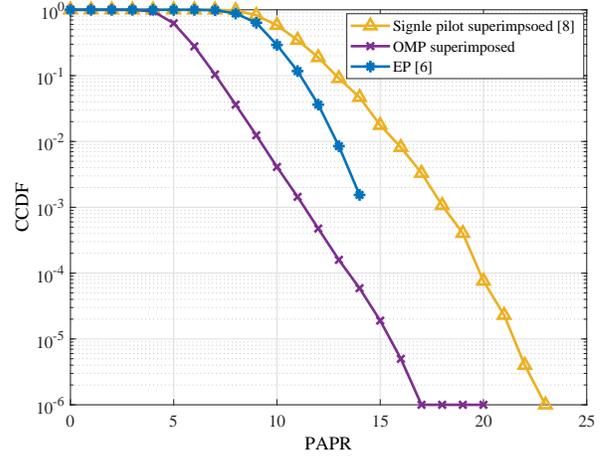}
	\caption{CCDF versus PAPR for the transmitted signal.}
	\label{papr}
\end{figure}

In this section, simulation results of the OMP superimposed channel estimation performance are illustrated and it is compared with the conventional EP scheme \cite{Raviteja2019}, the MMSE superimposed \cite{Mishra2021}, and the single pilot superimposed \cite{Yuan2021}. To evaluate the estimation performance, NMSE criterion is considered which is defined as NMSE$=||\mathbf{\hat{h}-h}||_2^2/||\mathbf{h}||_2^2$. Also, to compare the detection performance, the BER of the proposed method is compared with the ideal CSI and other estimation methods. Afterward, the PAPR of  the single pilot superimposed, the EP method, and the OMP superimposed are illustrated. The PAPR of the MMSE superimposed estimation is not shown as its PAPR is similar to the OMP superimposed. For a fair comparison, it is assumed that in all cases the total power of each frame is equal to one. It  means that each DD domain grid power must be equal to $\frac{1}{MN}$. Therefore, the pilot value for the EP scheme is $\sqrt{(l_{max}+1)(2k_{max}+1)}$ and the pilot value for the single pilot superimposed is $\sqrt{\sigma^2_pMN}$. Simulation parameters are set as follows. The gain and delay of channels tap are generated according to the one in \cite{Mishra2021}. The number of delays and Doppler grids are set to $M=16$ and $N=16$, respectively. 4-QAM modulation is used to transform input bits into symbols. Also, to generate Doppler for each tap the Jakes' formula is used which is fully explained in \cite{Raviteja2018a}.

Fig. \ref{nmmse} compares the NMMSE of different estimation methods. It shows that the OMP superimposed method can outperform the EP estimation for high SNRs. This is due to the fact that in high SNRs, in the first iteration, CSI estimation is done almost accurately. As the estimation error in the following iteration depends on the current iteration, the performance of the estimator is better than EP method in high SNRs. It is clear that in the low SNRs, since estimation in the first iteration does not estimate CSI accurately, the following iterations do not estimate CSI with low NMSE.  Also, Fig. \ref{nmmse} shows that the OMP superimposed performance is close to the MMSE superimposed. MMSE superimposed method and OMP superimposed method estimates the CSI more accurately than the single pilot superimposed. That is owing to the fact that in the OMP superimposed and  the MMSE superimposed method, channel estimation and data detection are employed jointly and information data helps the pilot to estimate CSI with more precision. It is clear that in the high SNRs for the superimposed estimations there is an error floor due to the interference from data to the pilot and it cannot be compensated completely.

Fig. \ref{ber} compares the BER curve versus SNR for the OMP superimposed with the EP estimation, the MMSE superimposed, and the single pilot superimposed. Also, the BER for the ideal CSI  is plotted for benchmark. It shows that the OMP superimposed method can reach the MMSE superimposed performance that is very close to the EP precision. It should be mentioned that our proposed method does not need any dedicated frame to estimate the delay and Doppler of each tap and is more spectrum efficient. Also, the MMSE estimation method assumes that the delay and Dopple of each tap is constant during a super-frame. Therefore, for highly TV channels, the overhead of the MMSE superimposed increases while this fact does not affect OMP superimposed. Moreover, the error floor is noticeable for high SNRs due to interference from the pilot to the received data. It was expected that the single pilot superimposed is not accurate enough.

Fig. \ref{papr} shows the CCDF of the PAPR for the OMP (or MMSE) superimposed, the EP, and the single pilot superimposed scheme. As the OMP superimposed scheme spreads pilot sequences in the DD domain, the probability of  high PAPR for the time domain signal is not high. However, for single pilot superimposed and EP scheme, one high-power pilot is located in the DD domain and this leads to high PAPR for the time domain signal.

\section{Conclusion}
In this paper, we proposed a superimposed channel estimation based on OMP algorithm and data detection using MP for the OTFS system. The proposed method did not impose pilot overhead to the system. Moreover, it did not need to dedicate an EP frame at the beginning of each super-frame. Also, its performance reached to the EP channel estimation. It should be mentioned that the proposed method had much less PAPR compared to EP and single superimposed pilot.


\bibliographystyle{IEEEtran}
\bibliography{source}

\end{document}